\documentclass[twocolumn,amsmath,amssymb,amsfonts,nobibnotes, nofootinbib,superscriptaddress,altaffilletter,aps,prd, reprint]{revtex4-2}

\usepackage[colorlinks=true,pdfstartview=FitV,linkcolor=blue,citecolor=blue,urlcolor=blue]{hyperref}
\usepackage{wasysym}
\usepackage[separate-uncertainty,retain-explicit-plus,per-mode=symbol,binary-units]{siunitx}
\usepackage{array,mathtools,amssymb,dcolumn}
\usepackage[below]{placeins}
\usepackage[dvipsnames, table]{xcolor}
\usepackage{tikz}
\usepackage{comment}
\usepackage{multirow}
\usepackage{afterpage}
\usepackage{physics}
\usepackage{paralist}
\usepackage{listings}
\usepackage{array}
\usepackage{url}
\usepackage{soul}
\usepackage{textcomp, gensymb}
\usepackage[english]{babel}
\usepackage{tabularx}
\usepackage{orcidlink}
\usepackage{tablefootnote}
\usepackage[font=small,figurename=FIG.,tablename=TABLE]{caption}
\newcommand{\TfftBin}{T_{\mathrm{FFT}}^{\mathrm{bin}}}
\newcommand{\Tfft}{T_{\mathrm{FFT}}}

\allowdisplaybreaks

\begin{document}
\setcounter{secnumdepth}{5}
\title{New semicoherent targeted search for continuous gravitational waves from pulsars in binary systems}

\newcommand{\UdSCag}{Physics Department, Universit\`a degli Studi di Cagliari, Cagliari  09042, Italy}
\newcommand{\INFNCA}{INFN, Sezione di Cagliari, Cagliari 09042, Italy}
\newcommand{\INFNRM}{INFN, Sezione di Roma, P.le A. Moro, 2, I-00185 Roma, Italy}
\newcommand{\Sapienza}{Physics Department, Universita’ di Roma “Sapienza”, P.le A. Moro, 2, I-00185 Roma, Italy}
\newcommand{\UCR}{Department of Physics and Astronomy, University of California, Riverside, CA 92521, USA}
\affiliation{\Sapienza}
\affiliation{\UdSCag}
\affiliation{\INFNRM}
\affiliation{\INFNCA}
\affiliation{\UCR}

\author{L.~Mirasola\,\orcidlink{0009-0004-0174-1377}}\altaffiliation{Contact author: \href{mailto:}{lorenzo.mirasola@ca.infn.it}}\affiliation{\UdSCag}\affiliation{\INFNCA}
\author{P.~Leaci\,\orcidlink{0000-0002-3997-5046}}\affiliation{\Sapienza}\affiliation{\INFNRM}
\author{P.~Astone\,\orcidlink{0000-0003-4981-4120}}\affiliation{\INFNRM}
\author{L.~D'Onofrio\,\orcidlink{0000-0001-9546-5959}}\affiliation{\INFNRM}
\author{S.~Dall'Osso\,\orcidlink{0000-0003-4366-8265}}\affiliation{\Sapienza}\affiliation{\INFNRM}
\author{A.~De~Falco\,\orcidlink{0000-0002-0830-4872}}\affiliation{\UdSCag}\affiliation{\INFNCA}
\author{M.~Lai\,\orcidlink{0000-0002-2388-9581}}\affiliation{\UCR}
\author{S.~Mastrogiovanni\,\orcidlink{0000-0003-1606-4183}}\affiliation{\INFNRM}
\author{C.~Palomba\,\orcidlink{0000-0002-4450-9883}}\affiliation{\INFNRM}
\author{A.~Riggio\,\orcidlink{0000-0002-6145-9224}}\affiliation{\UdSCag}\affiliation{\INFNCA}\affiliation{INAF/IASF Palermo, via Ugo La Malfa 153, I-90146 - Palermo, Italy}
\author{A.~Sanna\,\orcidlink{0000-0002-0118-2649}}\affiliation{\UdSCag}\affiliation{\INFNCA} 
\begin{abstract}
We present a novel semicoherent targeted search method for continuous gravitational waves (CWs) emitted by pulsars in binary systems. The method is based on a custom optimization of the coherence time, which is tailored according to the orbital parameters and their uncertainties, as provided by electromagnetic observations. While rotating pulsars are expected to produce quasimonochromatic CWs in their reference frame, their orbital motion introduces additional modulation in the observer’s frame, alongside the modulation caused by the Earth's motion. As a result, the received signal is spread across a frequency range, and demodulation techniques must be used to improve sensitivity.
However, Doppler corrections can, in some cases, vary significantly within the uncertainties of the orbital parameters, potentially lowering the detection chances of single-template, fully coherent searches. To exploit the constraints derived from electromagnetic observations, we implement a semicoherent search that is more robust than other methods. In this approach, the coherence time is evaluated for each source, taking into account the uncertainties in its orbital parameters. 
This method was tested and applied to a set of thirteen targets from the ATNF catalog. The search identified one outlier, whose astrophysical origin has been confidently excluded. For the first time to our knowledge, we then set upper limits on the signal strain from these 12 pulsars, with the lowest limit being $h_{UL}\sim9.01\times 10^{-26}$ for PSR~J1326-4728B.
\end{abstract}

\maketitle

\section{Introduction}
While $\mathcal{O}(10^8-10^9)$ neutron stars (NSs) are thought to exist in our Galaxy~\cite{Reed_2021}, only a few thousand have been observed thanks to their electromagnetic (EM) emission.~Most~of~these are rotation-powered radio pulsars whose emission beam~points along our line of sight, while a fraction is known from their broadband emission, which can be powered by accretion, dissipation of magnetic energy or even residual thermal energy~\cite{Hakobyan:2022kiy}.~Besides their EM emission, rotating NSs are expected to radiate continuous gravitational waves (CWs) if a shape asymmetry is present~\cite{Wette:2023dom}. CW emission is essentially monochromatic in the NS reference frame. It can be driven by elastic, thermal, or magnetic anisotropic stresses in the NS crust or core~\cite{PhysRevLett.95.211101}, or from a mass-current quadrupole induced by core fluid instabilities like, e.g., the $r$-modes~\cite{r-modes}.~Detecting CW signals will therefore allow us to determine the NS mass quadrupole, hence its degree of asymmetry, and eventually get insights on the NS interior structure as well as on the equation of state (EoS) of matter at supra-nuclear density (see, e.g.~\cite{Piccinni:2022vsd} and references herein).
\par All-sky searches for CWs represent to date the only means by which we can unveil the huge population of undetected NS in the Galaxy.~These all-sky searches for CWs have already been carried out in Advanced LIGO and Advanced Virgo data from previous science runs, leading to upper limits on the gravitational wave (GW) signal strain from NS in the galactic population both for isolated stars (e.g.,~\cite{PhysRevD.96.062002,PhysRevD.100.024004,O3_paper, PhysRevD.94.102002, Steltner_2021, Steltner_2023, PhysRevD.96.122004, PhysRevX.13.021020,PhysRevD.109.022007}), and in binary systems (e.g.,~\cite{Goetz_2011,PhysRevD.103.064017,PhysRevLett.124.191102, Covas_2022}).
\par On the other hand, the EM emission of known NSs, both isolated and in binary systems, provides accurate measurements of their sky location, rotational and, in case, binary orbital parameters. For these sources, searches have been performed looking for a corresponding CW signal (e.g.,~\cite{LIGOScientific:2021hvc,PhysRevD.105.022002,Leaci:2016oja, Whelan_2023,Abbott_2022_scoX1,PhysRevD.108.122002,Singhal_2019,PhysRevD.96.122006, PhysRevLett.120.031104, PhysRevD.106.042003, PhysRevD.95.082005, Ashok_2021, 10.1093/mnras/stac3742, Nieder_2020}), either using a single-template approach (targeted) or rather exploring a limited range around the known parameters (narrow-band or directed).
\par In this work, we present a novel approach to targeted searches of CWs from known pulsars in binary systems.~Uncertainties on the orbital parameters for such systems can sometimes affect the accuracy of Doppler corrections leading to a loss of signal-to-noise ratio (SNR)~\cite{PhysRevD.91.102003} if not considered.~To tackle this problem, we adopt a semicoherent technique in which data are first divided into segments, each of which is analyzed coherently before all segments are finally re-combined incoherently.~Semicoherent methods are less sensitive but more robust than single-segment analyses, as they do not rely on a strict phase-coherence of the signal over extended time spans (NSs CW emission can be affected by stochastic processes such as glitches~\cite{Ashton:2017wui} or accretion-induced spin wandering~\cite{Mukherjee:2017qme}). The novelty of our method is to choose a customized length of the segments according to the combined uncertainty of the neutron star binary orbital parameters.~In this way, we can perform a single-template search and reduce the potential significance loss, even though with some reduction in sensitivity.
\par The method has been tested and applied to a subset of pulsars from the ATNF catalogue\footnote{see~\cite{ATNF_link} for updated versions.} (version 1.70) using O3 data\footnote{Third observing run of LIGO-Virgo-KAGRA~\cite{ligo,virgo,kagra} detectors that took place between April 1 2019 and March 27 2020~\cite{PhysRevLett.123.231107,PhysRevLett.123.231108}.}. Although throughout this paper we use only our new semicoherent method, our coherence time calculation highlights a fraction of targets for which a fully coherent single-template approach could be safely applied without the need for a grid. Future searches can, in principle, make use of this criterion to decide whether a single or multi template search is needed for a given ephemeris.
\par The plan of the paper is as follows:~Sec.~\ref{sec:signal} describes the adopted phase model for the GW signal. Section~\ref{sec:pulsars} lists the selection criteria of the chosen pulsars in binary systems as extracted from the ATNF catalogue. Section~\ref{sec:method} outlines the analysis method, together with its validation. A thorough discussion of the results is provided in Sec.~\ref{sec:results}, which also presents the upper limit calculation.~Section~\ref{sec:conclusion} summarises our results and their relevance for future searches, as well as discusses possible future developments.  \section{The expected signal}
\label{sec:signal}

Here, we briefly describe the expected CW signal from a rotating NS, either isolated or in a binary system.

\subsection{Isolated neutron star phase}
The signal strain from a nonaxis-symmetric NS, which is rotating around one of its principal axes, can be expressed in the detector reference frame~\cite{Jaranowski:1999pd} as
    {\small\begin{align}
        \label{Eq:hSignal}
        h(t)  = h_0\,F_+(t;\vec{n},\Psi)\,  \frac{1+\cos^2\iota}{2}\, \cos \phi(t) +\nonumber\\
        h_0\,F_\times(t;\vec{n},\Psi)\, \cos\iota\,\sin \phi(t), 
    \end{align}}
where $\phi$ represents the GW phase (including modulations from the Earth's motion) and $\iota$ the inclination angle of the NS rotational axis with respect to the line of sight; $F_{+, \times}$ are the time-dependent beam pattern functions~\cite{Jaranowski:1999pd} that encode the detector response as a function of the source sky position $\vec{n}$, and of the polarization angle $\Psi$. The CW amplitude is defined for a quadrupolar deformation as
    {\small\begin{align}
        \label{Eq:CWstrain}
        h_0 = \frac{4\pi^2G}{c^4}\frac{I_{zz} \varepsilon  \,f_{\rm 0}^2\,}{d}, 
    \end{align}} 
where $\varepsilon = (I_{xx}-I_{yy})\>/\>I_{zz}$ is the equatorial ellipticity and $f_{\rm 0}$ the signal frequency, i.e., the time derivative of $\phi$ over $2\pi$ evaluated at a certain reference time $t_{\rm ref}$. For rotating nonaxis-symmetric NSs, $f_0$ is expected to be twice the rotational frequency, namely $f_{\rm 0}~=~2~f_{\rm rot}$.
\par Equation~\eqref{Eq:hSignal} can be rewritten as the real part of~\cite{Hplus_Hcross}
    {\small\begin{equation}
    \label{eq:hSignal_HplusHcross}
        h(t)  = H_0\, (A^+ H_+ + A^\times H_\times)\, e^{j\phi (t)},
    \end{equation}}
where $A^{+/\times} = F^{+/\times}(t;\Vec{n}, 0)$, and
    {\small\begin{align}
        H_0 = \frac{h_0}{2} \sqrt{1 + 6\, \cos^2\,\iota + \cos^4\,\iota}\,, \label{eq:Capital_H}\\
        H_{+} = \frac{\cos\,2\Psi\>- j\, \eta_{\rm pol} \,\sin\,2\Psi}{\sqrt{1+\eta_{\rm pol}^2}}\,,\label{eq:Hplus}\\
        H_{\times} = \frac{\sin\,2\Psi\>+ j\, \eta_{\rm pol} \,\cos\,2\Psi}{\sqrt{1+\eta_{\rm pol}^2}}\,,\label{eq:Hcross}
    \end{align}}
being
    {\small\begin{equation}
        \eta_{\rm pol} = -\frac{2\, \cos\,\iota}{1+\cos^2\,\iota}\, 
        \label{eq:eta_H0}
    \end{equation}}
the semiminor to semimajor axis ratio of the polarization ellipse.~In particular, $\eta_{\rm pol} = 0$ for a linearly polarized wave and $\eta_{\rm pol} = \pm 1$ for a circularly polarized wave, where the 
$+ (-) $ sign refers to the counterclockwise (clockwise) direction.
\par Moreover, the intrinsic signal frequency slowly decreases over time due to EM- and GW-driven rotational energy losses of the NS, which cause a secular spin-down\footnote{EM-driven spin-down is usually largely dominant in known pulsars.}.~The phase evolution can be expanded in the NS reference frame as
    {\small\begin{align}
        \label{Eq:dopp_mod}
        \phi^{\rm src}(\tau) = 2\pi\left[f_{\rm 0} (\tau - t_{\rm ref}) + \frac{\Dot{f}_{\rm 0}}{2} (\tau - t_{\rm ref})^2 + ... \right] + \phi_0,
    \end{align}}
where $\tau$ is the time at the source and $\phi_0$ the initial phase of the signal at $t_{\rm ref}$.~Since the Earth moves relative to the source, the induced Doppler modulation can be described as a temporal delay between the CW arrival time at the detector, $t_{\rm arr}$, and $\tau$.
\par In order to remove the Doppler modulation,~we must relate $\tau$ to $t_{\rm arr}$ such that $\phi(t_{\rm arr})= \phi^{\rm src}(\tau(t_{\rm arr}))$. Following Eq.~(5) of~\cite{Leaci:2016oja}, $\tau(t_{\rm arr})$ can be written for an isolated NS as
    {\small\begin{align}
        \label{Eq:source_proper_time}
        \tau(t_{\rm arr}) = t_{\rm arr} + \frac{\Vec{r}_{\rm SSB}\cdot \Vec{n}}{c} -  \Delta_E - \Delta_S\, - \frac{D}{c},
    \end{align}}
where the second term is the Earth's R\o{}mer delay, being $\Vec{r}_{\rm SSB}$ the vector from the Solar System barycenter (SSB) to the detector. It represents the vacuum delay between the arrival time at the observatory and the SSB. The third contribution is given by the Einstein delay ($\Delta_E$), which takes into account the redshift due to the other Solar System bodies. The term $\Delta_S$ is the Shapiro delay, linked to the deflection of a signal passing close to the Sun. The last term is the signal (light) vacuum travel time from the source, at a distance $D$, to the SSB. Following~\cite{Leaci:2016oja}, $D/c$ can be neglected by redefining the intrinsic spin-down parameters.

\subsection{Binary motion phase}
\label{subsect:binary_orbital_motion}
NSs in binary systems represent a more challenging target since their orbital motion induces an additional Doppler effect that must be accounted for in the analysis.~The orbital motion is parametrized as a time-dependent delay added to Eq.~\eqref{Eq:source_proper_time}, 
    {\small\begin{align}
        \tau_{\rm bin} = \tau(t_{\rm arr}) - \frac{R(\tau_{\rm bin})}{c} - \Delta_E^{\rm bin} - \Delta_S^{\rm bin}\,,
    \end{align}}
where $\Delta_E^{\rm bin}$ and $\Delta_S^{\rm bin}$ are, respectively, the Einstein and Shapiro delay due to the binary companion and to possible planets in the system~\cite{Tempo2_1,Tempo2_2}.~Finally, $R/c$ represents the R\o{}mer delay of the binary system, with $R(\tau_{\rm bin})$ being the radial distance of the CW-emitting NS from the binary barycenter (BB), projected along the line of sight.
\par Following the discussion in~\cite{Leaci:2016oja,PhysRevD.91.102003}, $R/c$ can be expressed in terms of five binary orbital parameters: $a_p$, the projected semimajor axis of the orbit along the line of sight normalized to $c$; $P_{\rm orb}$, the orbital period; $e$, the orbit eccentricity; $\omega$, the argument of periastron; and $t_p$, the time of periapse's passage. Therefore, the binary delay is given by
    {\small\begin{align}
        \label{eq:R_c_high_ecc}
        \frac{R}{c} = a_p\>\left[\sin\omega\>(\cos E-e)+\cos\omega\>\sin E\>\sqrt{1-e^2}\right]\,,
    \end{align}}
where $E$ is the eccentric anomaly, expressed through Kepler's equation
    {\small\begin{align}
        \tau_{\rm bin} - t_p = \frac{P_{\rm orb}}{2\pi}\>(E-e\>\sin E)\,.
    \end{align}}
In the regime $a_p\ll P_{\rm orb}$, which is normally the case for the observed systems, we can use the additional approximation of $E(\tau_{\rm bin})\approx E(t)$ as the change in $E$ during the time $R/c$ will be negligible. As a consequence, defining $t\equiv~\tau(t_{\rm arr})$, $E(t)$ becomes
    {\small\begin{equation}
    \label{eq:kepler_eq}
        t - t_p \approx \frac{P_{\rm orb}}{2\pi}\>(E-e\>\sin E)\,.
    \end{equation}}

Moreover, since eccentricity tends to be dissipated with time~\cite{PhysRev.136.B1224}, it is useful for relatively old binary systems to expand Eq.~\eqref{eq:R_c_high_ecc} in the small eccentricity regime up to the leading order in $e$. The binary R\o{}mer delay can be then expressed as~\cite{PhysRevD.91.102003}
   {\small\begin{align}
        \label{eq:R_c_low_ecc}
        \frac{R}{c} \approx a_p \left[ \sin\psi(t) + \frac{\kappa}{2}\sin 2\psi(t) - \frac{\eta}{2}\cos 2\psi(t) -\,\frac{3}{2}\eta\right]\,,
   \end{align}}
where $\eta\,, \kappa$ represent the \textit{Laplace-Lagrange} parameters, and $\psi(t)$ is the mean anomaly measured with respect to the time of ascending node $t_{\rm asc}$. Defining the mean orbital angular velocity $\Omega = 2\pi/P_{\rm orb}$, the mentioned parameters can be expressed as
    {\small\begin{align}
        \kappa= e\cdot\cos\omega\,,\quad  \eta = e\cdot\sin\omega\,,\label{eq:kappa-eta}\\
        \psi (t) = \Omega\, (t - t_{\rm asc})\,, \quad t_{\rm asc} = t_p - \frac{\omega}{\Omega}\label{eq:psi-tasc}\,.
    \end{align}} \section{Target selection\label{sec:pulsars}}
Our analysis was carried out on a subset of pulsars in binary systems from version 1.70 of the ATNF catalog~\cite{Manchester_2005}.~We selected targets for which the NS spin period reference time was within the O3 run and for which all binary orbital parameters are measured. The first selection criterion is justified to avoid less precise and potentially incorrect extrapolated values of the source parameters. A list of our selected targets and their parameters of interest is given in Table~\ref{tab:candidates}.
\par The ephemerides are obtained by fitting the time of arrival (ToA) of photons emitted by a pulsar. To do so, astronomers exploit a time model that, in the case of pulsars in binary systems, is constituted either by ($a_p,\, P_{\mathrm{orb}},\,e,\,\omega,\,t_p$), or ($a_p,\, P_{\mathrm{orb}},\,\eta,\,\kappa,\,t_{\mathrm{asc}}$) when the eccentricity is small\footnote{For reproducibility, we refer that the ATNF ephemerides mistakenly reported some parameters. We found, in the reference paper \cite{MNRAS.518.1672M}, seven pulsars for which the ATNF $t_p$ was, instead, the $t_\mathrm{asc}$: J1811-0624, J1813-0402, J1824-0621, J1912-0952, J1929+0132, J2001+0701, J2015+0756. While for an eighth pulsar, J2338+4818, the $t_p$ value was reported as $t_\mathrm{asc}$ \cite{10.1093/mnras/stab2540}.}.

\begin{table*}[htb]
\begin{center}
\caption{List of selected targets from the 1.70 ATNF catalog version. The reference epoch for the spin frequencies is set to the Modified Julian Date (MJD) 58754.5. A description of the binary orbital parameters is given in Sec.~\ref{subsect:binary_orbital_motion}. The last column shows the upper limits at the $95\%$ confidence level on the strain amplitude for each target, see Sec.~\ref{sec:results} for details.}
 \renewcommand*{\arraystretch}{1.4}
 \resizebox{\textwidth}{!}{\begin{tabular}{cccccccc} 
 \hline\hline 
  Name & $P_{orb}$ [day] & $a_p$ [lt-s] & $e \,/\, \eta$ & $\omega$ [deg] / $\kappa$ & $t_{\mathrm{p}} / t_{\mathrm{asc}}$ [MJD] & $f_{\rm 0}$ [Hz] & $h_{\rm UL}$\\ \hline
J1326-4728B & 0.08961121(1)  &	0.021455(7) & 1(6)$\cdot10^{-4}$$^*$ & -4(6)$\cdot10^{-4}$$^*$ & 58768.037243(4)$^*$ & 417.37366625(1) & 8.1$\cdot10^{-26}$\\\hline
J1701-3006G & 0.77443355(5)  &	0.620316(8) & 6.1(3)$\cdot10^{-4}$$^*$ & 6.7(3)$\cdot10^{-4}$$^*$ & 58894.966321(3)$^*$ & 434.0181102(12) & 7.3$\cdot10^{-26}$\\\hline
J1801-0857B & 59.836454(1)   & 33.87543(1)  & 3.82258(7)$\cdot10^{-2}$ & 302.1086(6) & 54757.72304(9) & 69.0569857338(2) & 1.2$\cdot10^{-25}$\\\hline
J1811-0624  & 9.38782916(5)  &	6.561124(2) & -3.8(5)$\cdot10^{-6}$$^*$ & -2.61(12)$\cdot10^{-5}$$^*$ & 58692.1778781(8)$^*$& 482.429490052(2) & 1.2$\cdot10^{-25}$\\\hline
J1813-0402  & 10.560352898(4)& 	9.4263516(5)& -2.91(17)$\cdot10^{-6}$$^*$ & -2.97(12)$\cdot10^{-6}$$^*$	& 58716.88174915(15)$^*$ & 487.02569177524(4) & 1.3$\cdot10^{-25}$\\\hline
J1823-3021G & 1.54013654(5)  &	3.00331(2)  & 3.80466(6)$\cdot10^{-1}$  & 146.721(1)  & 58909.681498(8)  & 328.337491175(2) & 9.3$\cdot10^{-26}$\\\hline
J1824-0621  & 100.91365361(5)& 	44.103911(2)& -22.809(2)$\cdot10^{-5}$$^*$ & -2.363(2)$\cdot10^{-5}$$^*$ & 58734.9698712(18)$^*$ & 618.61737095733(4) & 1.13$\cdot10^{-25}$\\\hline
J1835-3259B & 1.19786323(1)  &	1.430841(6) & -2.7(4)$\cdot10^{-5}$$^*$ & -2.2(3)$\cdot10^{-5}$$^*$ & 58866.2655109(6)$^*$  & 1092.7211722859(2) & 1.6$\cdot10^{-25}$\\\hline
J1912-0952  & 29.4925321(5)  &	5.379447(7) & -1.62(18)$\cdot10^{-5}$$^*$  & 4.6(3)$\cdot10^{-5}$$^*$ & 58685.963815(4)$^*$ & 79.78283667171(8) & 1.3$\cdot10^{-25}$\\\hline
J1929+0132  & 9.265037044(4) &	8.2675144(6)& -8.12(18)$\cdot10^{-6}$$^*$ & -2.44(18)$\cdot10^{-6}$$^*$	& 58757.4798650(2)$^*$ &311.61302742043(4) & 6.7$\cdot10^{-26}$\\\hline
J2001+0701  & 5.22708263(2)  & 	2.23028000(15)& -1.85(12)$\cdot10^{-5}$$^*$ & 1.29(12)$\cdot10^{-5}$$^*$ & 58754.681385(11)$^*$ & 293.25691384168(5) & 9.0$\cdot10^{-29}$\\\hline
J2015+0756  & 6.456719142(2) &	5.1642574(2)  &	-1.3(1)$\cdot10^{-6}$$^*$ & 2.80(13)$\cdot10^{-6}$$^*$   & 58780.4740740(2)$^*$ & 462.10892181919(18) & 9.9$\cdot10^{-26}$\\\hline
J2338+4818  & 95.25536(2) &	117.58572(7) &	1.8237(9)$\cdot10^{-3}$ & 99.65(2)   & 58868.444(7) & 16.84774472985(17) & 1.2$\cdot10^{-23}$\\\hline
 \hline
\end{tabular}}
 \label{tab:candidates}
 {\small{$^*$These values refer to the low-eccentricity parameters $\eta$, $\kappa$ and $t_{\rm asc}$, see Eqs.~\eqref{eq:kappa-eta}-\eqref{eq:psi-tasc}, used during the analysis.}}
\end{center}
\end{table*}

 \section{Analysis method}
\label{sec:method}

The binary correction can strongly depend on the orbital parameters, implying that a slight offset from the true values, e.g. due to uncertainties on their measured values, can lead to a substantial loss of phase-coherence, hence SNR (see, e.g.,~\cite{PhysRevD.91.102003}). For this reason, a semicoherent approach can overcome the potential computational burden of a multitemplate search to cover the parameter-space region of uncertainties.
However, semicoherent methods are less sensitive with respect to fully coherent searches (see Eqs.~(72-73) from~\cite{Astone:2014esa}) as
    {\small\begin{equation}
        \frac{h_{\rm sens}^{\rm (coh)}}{h_{\rm sens}^{\rm (semicoh)}} \simeq \left(\frac{\Tfft}{T_{\rm obs}}\right)^{1/4}
        \label{eq:sens_ratio}
    \end{equation}}
which is $\sim 23\%$ for $T_{\rm obs} = 1$~yr and $\Tfft = 1$~d. This rough estimate does not consider further parameters, such as the number of selected outliers and the adopted detection statistics, upon which the sensitivity also depends (see e.g.~\cite{Astone:2014esa}).
Here, we determine the maximal timespan over which phase coherence can be maintained, i.e. a coherence time, based on the uncertainties of the measured binary orbital parameters. 

\par We propose a semicoherent search method which makes use of calibrated and cleaned detector data preprocessed through the \textit{band sampled data}~\cite{Piccinni:2018akm} (BSD) framework. The method is structured as follows: for each target, we first perform a Doppler correction in the time domain with the ephemeris central values, and then we divide the data into half-interlaced chunks of duration $\Tfft$, the so-called coherence time:~its evaluation, based on the binary parameters and their uncertainties, is a crucial step of this analysis that we detail in Sec.~\ref{sec:coherence_time_eval}.\\
Each chunk is windowed to reduce the finite length effect and then Fourier transformed.~The next step is the selection of peaks from the equalized spectrum\footnote{The equalized spectrum is the ratio between the squared modulus of the Fourier-transformed chunk and the average single-sided \textit{power spectral density}~\cite{Piccinni:2018akm}.} of the Fourier-transformed chunk~\cite{Astone_2005,Astone:2014esa}: a frequency bin of this spectrum is selected only if it is a local maximum and its height exceeds a threshold $\theta_{\rm thr}=2.5$. As demonstrated in~\cite{Astone:2014esa}, $\theta_{\rm thr}$ is linked to the false alarm probability (FAP) per chunk $p_0$ through the relation 
    {\small\begin{equation}
        p_0 = e^{-\theta_{\rm thr}} - e^{-2\theta_{\rm thr}}+ \frac{1}{3} e^{-3\theta_{\rm thr}}\,,
        \label{eq:FAP_single_Tfft}
    \end{equation}}
that, for our threshold, corresponds to a value of $\approx 0.076$.\\
After that, those selected peaks along the observing run are collected together in a time-frequency map that we call \textit{peakmap}.

    \begin{figure*}[btp]
        \includegraphics[scale=0.34]{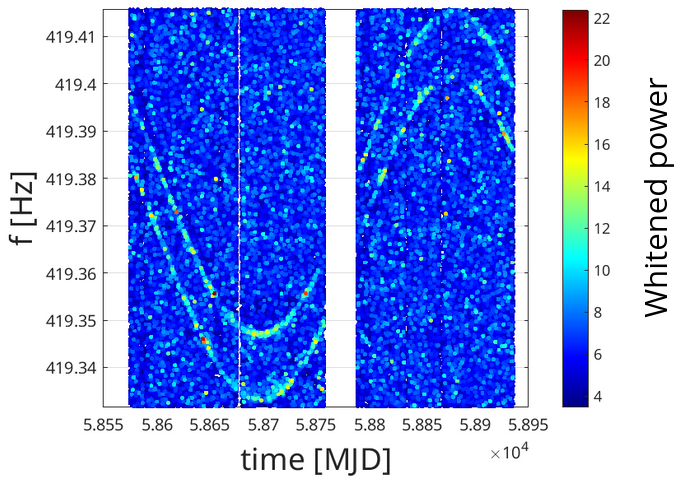}
        \includegraphics[scale=0.34]{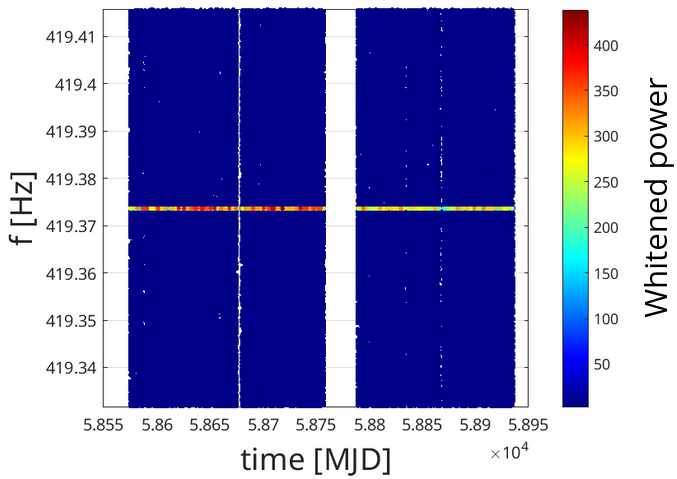}
        \caption{Peakmaps of a fake signal injected with $h_0 = 10^{-24}$ into real LL data from the O3 run over a timespan of $\sim 1$~yr using $\Tfft = T_{\rm sid}/5$, without (\textit{left}) and with (\textit{right}) the Doppler correction. The signal mimics that expected from J1326-4728B, i.e., with the same binary orbital parameters and the same sky position, while the injected frequency is $f_{GW} = 2f_{rot}+2$~Hz to avoid the overlap with a possible true signal.}
        \label{fig:PM_before_after_Doppl}
    \end{figure*}

\par The peakmap is then projected onto the frequency axis and the loudest peak around the expected frequency, in the range $f_{\rm 0}\pm 1.5\> \delta f$\footnote{The choice of 1.5 bins around $f_{\rm 0}$ is done after systematics studies with injections.} is selected as a \textit{candidate}, where $\delta f$ is the Fourier transform frequency resolution given by
    {\small\begin{equation}
        \label{eq:delta_f}
        \delta f = \frac{1}{\Tfft}\,.
    \end{equation}}
    
Then the \textit{critical ratio} (CR), our detection statistics~\cite{Astone:2014esa}, is evaluated for the chosen candidate as
   {\small\begin{equation}
        \label{eq:CR}
        CR = \frac{x-\mu}{\sigma}\,,
    \end{equation}}
where $x$ is the candidate's number of counts in the projection of the peakmap (i.e., the number of times in which that specific bin has been selected), $\mu$ and $\sigma$ are, respectively, the peakmap projection's average and standard deviation estimated outside the region from which the loudest peak is selected. The CR is then used to assess the statistical relevance of the candidate. As discussed in~\cite{Astone:2014esa}, $\theta_{\rm thr}=2.5$ is a typical value used in all-sky searches that represents a compromise among a small sensitivity loss ($\sim10\%$), maximizing the $CR$ in the presence of a signal and being robust against disturbances.
\par An example of a peakmap before and after applying the Doppler correction is shown in Fig.~\ref{fig:PM_before_after_Doppl}, where a Software Injection (SI, i.e., simulated signal) was performed in LIGO Livingston (LL) O3 real data.

\subsection{Coherence time evaluation \label{sec:coherence_time_eval}}
    
\par The frequency shift due to the orbital motion can be linked to the time derivative of Eqs.~(\ref{eq:R_c_high_ecc}-\ref{eq:R_c_low_ecc}) as~\cite{Leaci:2016oja}
    {\small\begin{equation}
        \frac{\Delta f_{GW}^{\rm Doppl}}{f_{GW}}(t) = \frac{\Dot{R}}{c}(t) \,,
        \label{eq:doppl_shift_bin}
    \end{equation}}
where $\Delta f_{GW}^{\rm Doppl}$ is the time-dependent difference between the binary modulated frequency (i.e., the signal frequency after removing only the Doppler modulation due to the source's sky location) and $f_{\rm GW} = f_0 + \Dot{f}_0 (t-t_{\rm ref})$. On the other hand, the right-hand side term is
  {\small\begin{equation}
  \label{eq:high_ecc_R_c_dot}
        \frac{\Dot{R}}{c} = a_p \, \Omega \, \displaystyle \frac{\cos\omega\,\cos E\,\sqrt{1-e^2}-\sin\omega\,\sin E}{1-e\,\cos E}\,,
  \end{equation}}
where $E$ is evaluated with Eq.~\eqref{eq:kepler_eq}, and in the limit of low eccentricities, Eq.~\eqref{eq:high_ecc_R_c_dot} can be expressed as~\cite{PhysRevD.91.102003}
    {\small\begin{equation}
    \label{eq:low_ecc_R_c_dot}
        \frac{\Dot{R}}{c} \approx a_p \, \Omega \, \left(\cos\psi+\kappa\cos2\psi+\eta \sin2\psi\,\right) \,.
    \end{equation}}
We do not consider time derivatives of the orbital parameters as their variations are typically negligible over the characteristic duration of observing runs, $T_{\rm obs}\sim 1$~yr\footnote{In Eq.~\eqref{eq:doppl_shift_bin} we omitted the binary Einstein and Shapiro contribution, although they are used to demodulate the data.}.
\par We evaluate the coherence time of the targets, $\TfftBin$, requiring that the maximum\footnote{Selected along the observation time.} frequency shift caused by uncertainties on the Doppler correction (i.e., the variation of $\Delta f_{GW}^{\rm Doppl}$ with respect to the parameters) must be confined in half of a single frequency bin.\\
We calculate the residual Doppler modulation as
    {\small\begin{equation}
        \label{eq:delta_doppl}
        \delta\Delta f_{GW}^{\rm Doppl} (t) \simeq f_{GW}\cdot\sum_{i=1}^5\> \left.\frac{\partial \Dot{R}/c}{\partial x_i}\>\sigma_{x_i}\right\vert_{x_i=x_0}\,,
    \end{equation}}
where we omitted the time dependence of the right-hand side for simplicity. $x_i$ are the orbital parameters (see Sec.~\ref{sec:pulsars} and Table~\ref{tab:candidates}), $\sigma_{x_i}$ their uncertainties, and $x_0$ their central values as given in the ephemerides. For the sake of completeness, we detail the calculation of Eq.~\eqref{eq:delta_doppl} in Appendix~\ref{app:doppl_var}.\\
The coherence time of each source can then be estimated, assuming a dominant contribution from the orbital parameters only (see Appendix~\ref{app:Tfft_err_sky_pos}), as
    {\small\begin{equation}
        {\rm max}\left(\abs{\delta\Delta f_{GW}^{\rm Doppl}(t)}\right)_{t\in T_{\rm obs}} \leq \frac{\delta f}{2}\,,
        \label{eq:condition_coherence_time}
    \end{equation}}
where $\delta f$ is given in Eq.~\eqref{eq:delta_f}. We neglect the error on frequencies that are typically well constrained (see Table~\ref{tab:candidates}).   
\begin{figure*}[htpb]
        \centering
        \includegraphics[width=0.4\textwidth]{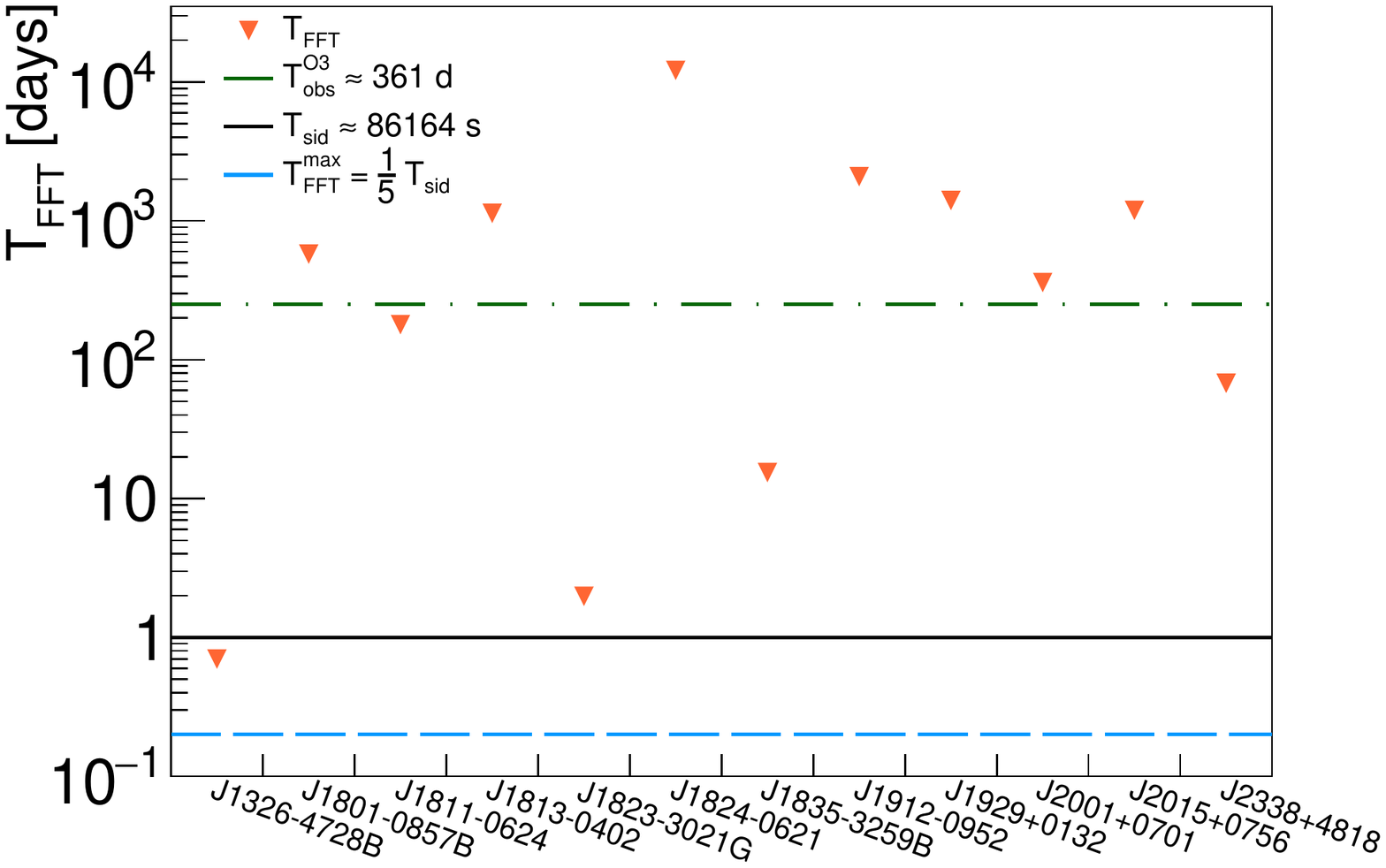}
        \includegraphics[width=0.4\textwidth]{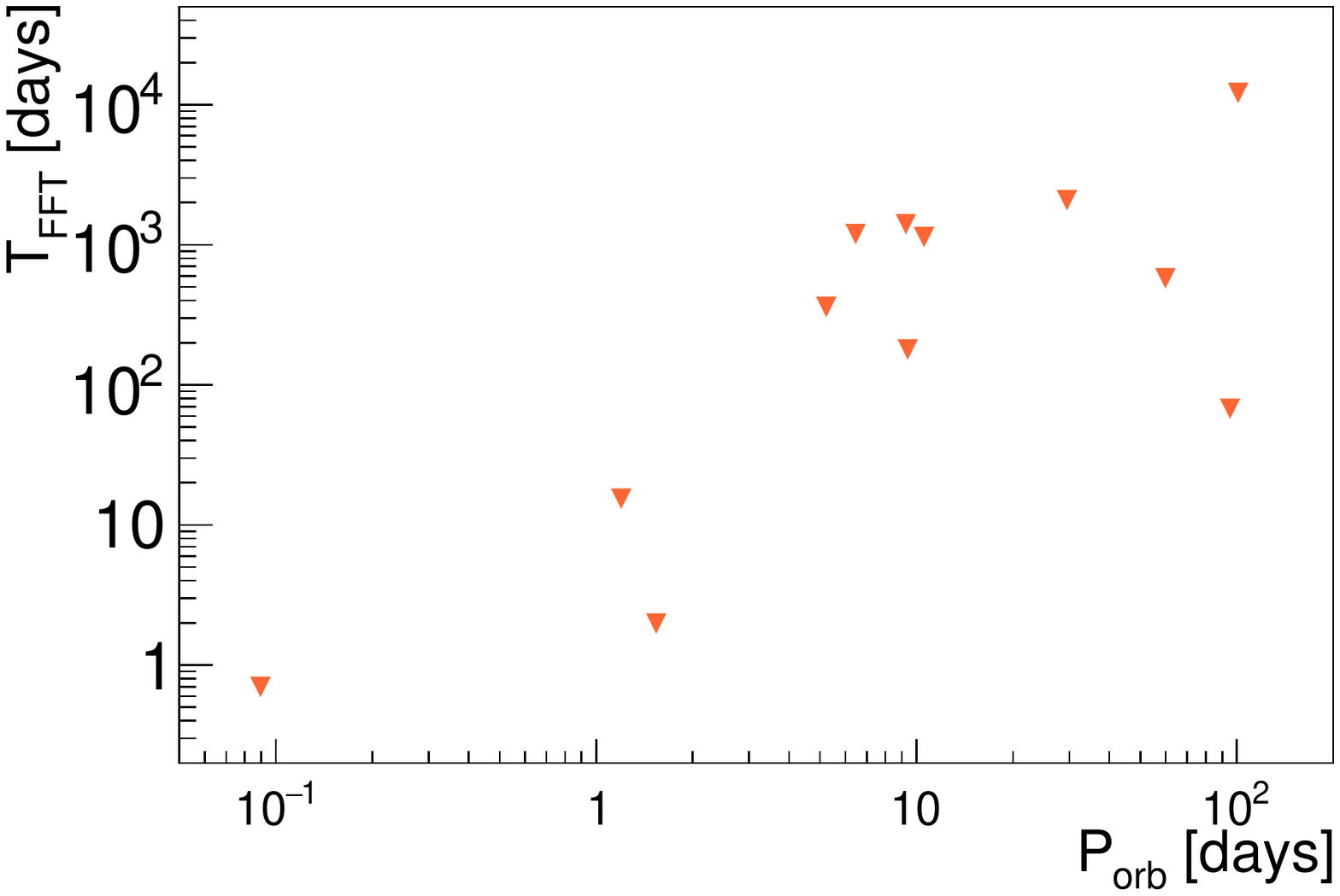}
        \caption{Left: Coherence time of each of the targets reported in Table~\ref{tab:candidates} according to Eq.~\eqref{eq:condition_coherence_time}. The sidereal day duration as well as the O3 duration are reported for comparison as the solid black and dashed-dotted green lines, respectively. The dashed light blue line is the upper value of $\frac{1}{5}\cdot T_{sid}$ used in this work to avoid the five peaks splitting, see text for details. Right: Coherence times plotted against the orbital period of the targets.}
        \label{fig:Tfft_targets}
    \end{figure*}

\par Finally, in the left plot of Fig.~\ref{fig:Tfft_targets}, we report the $\TfftBin$ of the targets in Table~\ref{tab:candidates} calculated with Eq.~\eqref{eq:condition_coherence_time}. We do not evaluate $\delta\Delta f_{GW}^{\rm Doppl} (t)$ using a squared sum as it would not capture potential anticorrelations among the orbital parameters, conservatively lowering the coherence time. For those targets for which $e,\,\omega$ are given (see Table~\ref{tab:candidates}), our estimation with Eq.~\eqref{eq:delta_doppl} is higher up to a factor of 5 than the squared sum calculation, being less conservative and potentially more sensitive. This choice is supported by the tests in Sec.~\ref{subsec:test_with_SI} that did not highlight strong SNR losses linked to the offset in the orbital parameters. On the other hand, when the low eccentricity parameters are given, our coherence time estimation is up to a factor 1.8 lower than the squared sum ones, meaning that we are more conservative in that scenario at the price of sensitivity loss (see Eq.~\eqref{eq:sens_ratio}).

Although in this work we describe and apply a semicoherent approach, the coherence time calculation shows that a fraction of targets can be safely studied with a single-template fully coherent approach as $\TfftBin>T_{\rm obs}^{O3}\sim1$~yr.
\par However, at present our method is bounded to $\Tfft$ shorter than about one sidereal day\footnote{The sidereal day represents the time taken by Earth to complete a full rotation around its rotational axis with respect to the stars. $T_{\rm sid}\sim 86164$~s.}. When $\Tfft$ approaches $T_{\rm sid}$, a CW monochromatic signal is split into five peaks at $f_{0} \pm k/T_{\rm sid}$, with $k=0,\>1,\>2$, due to the harmonic components of the beam pattern function~\cite{Jaranowski:1999pd}, with a significant loss in the sensitivity if not properly accounted for. 
Empirically and in agreement with generic all-sky searches~\cite{O3_paper}, we observe that if $\Tfft>T_{\rm sid}/5$ (see Appendix~\ref{app:Tfft_max}) we become sensitive to the spreading of signal power due to the Earth's motion and, consequently, encounter a loss in signal detection probability (i.e., sensitivity). For this reason, the present analysis is conservatively performed with a maximum coherence time equal to
    {\small\begin{equation}
        \Tfft^{\rm max}=\frac{1}{5}\cdot T_{\rm sid}\approx 17230 \text{ s,}
    \end{equation}}
for all our targets, since $\TfftBin>\Tfft^{\rm max}$ (see Fig.~\ref{fig:Tfft_targets}), without the need to take into account the splitting linked to the sidereal modulation. In Appendix~\ref{app:Tfft_ATNF_binaries} we highlight a population of potential targets with $\TfftBin<\Tfft^{\rm max}$ in the ATNF catalog that can benefit from this method.

We foresee overcoming this limitation by implementing in the semicoherent method discussed in~\cite{DAntonio:2023jxm} the binary Doppler correction that is now missing. This would allow for longer coherence times and better sensitivities.

\par We stress that, although we could not reach the estimated coherence times, shorter $\Tfft$ are always more reliable against offsets in the orbital parameter and stochastic processes such as glitches~\cite{Ashton:2017wui} or accretion-induced spin-wandering~\cite{Mukherjee:2017qme}. In that sense, our method is complementary to fully coherent ones that exploit a grid~\cite{Singhal_2019} or not~\cite{PhysRevD.108.122002} around the measured parameters, and represents an additional choice to other semicoherent pipelines such as~\cite{PhysRevD.105.022002}. 
Moreover, a major advantage of our method is that allows us to perform a single Doppler correction at the ephemeris central parameters. This is computationally very convenient compared with other pipelines~\cite{PhysRevD.105.022002}, where a grid around the estimated values is built to probe the parameter space. Indeed, if $\sigma_{par}$ (i.e., the uncertainty on a binary orbital parameter) is taken as a grid step, an interval of $\pm n_{\sigma}\sigma_{par}$ around each parameter becomes challenging even for small $n_\sigma$, since in each dimension we would explore in the worst case $(2n_{\sigma} +1)$ values. Hence, since the dimensionality of the binary parameter space is 5, the total number of points would be $N_{grid} = (2n_{\sigma} +1)^5$ values. Finer grids will be even more computationally challenging to scale toward larger pulsar datasets.

\subsection{Sensitivity estimation}
Following the conservative sensitivity estimation of~\cite{Astone:2014esa}, we can evaluate the expected minimum detectable strain (at a given confidence level $\Gamma$, CL), which would produce a candidate in a fraction $\geq \Gamma$ of a large number of repeated experiments, as
        {\small\begin{gather}
            h_{\rm sens}\approx
            \frac{4.02}{N^{1/4}_{FFT}\varepsilon^{1/4}\theta^{1/2}_{thr}}\sqrt{\frac{S_n(f)}{\Tfft}}\left(\frac{p_0(1-p_0)}{p^2_1}\right)^{1/4}\nonumber\\ 
            \times\sqrt{CR_{thr}-\sqrt{2} \mathrm{erfc}^{-1}(2\Gamma)} \propto \left(\frac{1}{T_{\rm obs}\Tfft}\right) ^{1/4}\,,
        \label{eq:h0min}
        \end{gather}}
where $S_n(f)$ is the unilateral noise spectral density, 
    {\small\begin{equation}
        p_1=e^{-\theta_{\rm thr}}-2e^{-2\theta_{\rm thr}}+e^{-3\theta_{\rm thr}}\,,
    \end{equation}}
$p_0$ is given in Eq.~\eqref{eq:FAP_single_Tfft}, $\varepsilon$ is the duty cycle of the considered detector~\cite{KAGRA:2023pio}, and $\mathrm{erfc}(x)$ is the complementary error function defined as~$\mathrm{erfc}(x)=$~$\frac{2}{\sqrt{\pi}}\int_x^{+\infty}{e^{-t^2}\,dt}$. The curves presented throughout this paper consider $CR_{thr}\approx2.79$, which comes from a chosen FAP in the case of white noise only of the $1\%$~(see Appendix~\ref{app:noise_distr}), $\theta_{thr}=2.5$~\cite{Astone:2014esa}, $\Gamma=95\%$, and the number of chunks $N_{FFT}$ is given by
    {\small\begin{equation}
        \label{eq:Nfft}
        N_{\rm FFT} = \frac{T_{\rm obs}}{\Tfft}\,.
    \end{equation}}
\par Since $h_{\rm sens}$ is proportional to $\left(T_{\rm obs}\cdot \Tfft\right) ^{-1/4}$, the higher the $\Tfft$, the better is the sensitivity. Hence, longer coherence times are preferred to shorter ones.

\subsection{Follow-up of potential outliers}
Outliers are identified when their $CR$ exceeds the threshold $CR_{thr}$.
In such circumstances, the peakmap is checked to discard potential correlation with a terrestrial noise. At the same time, we examine a list of known instrumental disturbances~\cite{O3Linelist} to verify that no overlap between the two is present.

    \begin{figure*}[htp]
        \centering
        \includegraphics[width=0.4\textwidth]{CR_wrong_corrections_err_nominal_Tfft_correct_ASD_correct_i_90_psi_30.pdf}
        \includegraphics[width=0.4\textwidth]{CR_wrong_corrections_err_smaller_Tfft_correct_ASD_correct_i_90_psi_30.pdf}
        \includegraphics[width=0.4\textwidth]{CR_wrong_corrections_err_larger_Tfft_correct_ASD_correct_i_90_psi_30.pdf}
        \includegraphics[width=0.4\textwidth]{CR_wrong_corrections_err_larger_Tfft_wrong_ASD_correct_i_90_psi_30.pdf}
        \caption{$r(d)$ [Eq.~\eqref{eq:r_of_d}] as a function of $d$ [Eq.~\eqref{eq:eucl_dist}] calculated with injections for pulsar J1823-3021G in simulated Gaussian noise data of $T_{\rm obs} = 1$~yr evaluated for the nominal uncertainties (top left), for smaller error (top right) and larger (bottom row) by a factor of 20 with respect to Table~\ref{tab:candidates}. $\TfftBin$ is the coherence time calculated with Eq.~\eqref{eq:condition_coherence_time}, while $\Tfft$ is the actual coherence time used in the analysis. Each dot represents a Doppler correction performed at a distance $d$, the shaded region denotes a fluctuation of one (darker blue) and two (light blue) $\sigma$ of the noise in the projection of the peakmap. See text for details.}
        \label{fig:CR_vs_dist}
    \end{figure*}    
    
In the scenario where these tests are passed, we identify a potential pipeline described in~\cite{PhysRevD.97.103020,PhysRevD.104.084012} to follow up the candidates while taking care of uncertainties in the orbital parameters.
This method has the flexibility to be used in both fully or semicoherent mode and can take care of all the uncertainties as well assuming the same signal model presented in Sec.~\ref{sec:signal}. It is based on a Monte Carlo Markov Chain (MCMC), where the template bank is not fixed by a grid but rather randomly sampled according to a posterior probability. In a recent work~\cite{Mirasola:2024lcq}, important warnings and features of this pipeline have been highlighted.
Spin-wandering and unknown glitches effects would be tackled by gradually increasing the coherence time, and checking the evolution of the reconstructed SNR. Since these effects are not modelled yet, we can follow~\cite{Mirasola:2024lcq} by stopping the follow-up after a certain coherence time.
Conclusive statements should be made upon the dataset at hand.
\subsection{Test on injected signals}\label{subsec:test_with_SI}    
The computational efficiency of our new method relies on a single Doppler correction applied to each target, without the need to scan across a grid in parameter space set by the uncertainties on orbital parameters. 
    
As a test, we simulate gapless Gaussian noise data with $T_{\rm obs} = 365$~d and $\sqrt{S_n} = 10^{-23}/\sqrt{\mathrm{Hz}}$ with a location consistent with~LL. We then inject a set of signals that mimics the emission of the targets in Table~\ref{tab:candidates}.
\par Each SI is injected with fixed $h_0$ calculated with Eq.~\eqref{eq:h0min} for $CR=5$ and $\Gamma = 95\%$, then we use $\cos\iota = 1$ and $\Psi = \ang{30}$ to circumscribe our test to the orbital parameters only. Additional tests to verify the independence on $\cos\iota$ and $\Psi$ are shown in Appendix~\ref{app:other_par_test}.
\par The validity is estimated employing the ratio $r$ between the critical ratio obtained with the exact Doppler correction (on target, OT) and when an offset $d$ is introduced, namely
{\small\begin{equation}
    r(d) = \frac{CR_{\rm OFF}}{CR_{\rm OT}}\,,
    \label{eq:r_of_d}
\end{equation}}
where $d$ is
{\small\begin{equation}
    d = \sqrt{\sum_{i=1}^5 \left(\frac{x_i^{\rm corr} - x_i^{\rm inj}}{\sigma_{x_i}}\right)^2}\,,
    \label{eq:eucl_dist}
\end{equation}}
and represents the Euclidian distance between the parameters used for the correction and the injected ones, normalized to their uncertainties.
\par In the test, we calculate $r(d)$ for each target varying the parameters by adding to the central value $[\pm 1,\pm 2] \sigma_{x_i}$\footnote{A smaller offset is expected in the real scenario since it is unlucky to have all the real parameters outside a $\pm \sigma_{x_i}$ interval from the measured values.}. 

As expected, the uncertainties are not playing a central role (see top left plot in Fig.~\ref{fig:CR_vs_dist} for J1823-3021G) due to the fact $\Tfft < \TfftBin$ (see Fig.~\ref{fig:Tfft_targets}).

\par However, we made a further test on a subset of three pulsars (J1326-4728B, J1701-3006G, and J1823-3021G) as an additional step of the method's validation. Here, we artificially vary the uncertainties given in the ephemerides to test different regimes:
\begin{itemize}
    \item[(i)] $\TfftBin\gg \Tfft^{max}$,
    \item[(ii)] $\TfftBin \sim \Tfft \ll \Tfft^{max}$,
    \item[(iii)] $\TfftBin < \Tfft$,
\end{itemize}
where we shrink/enlarge the uncertainties by a factor of 20. For simplicity, we show in Fig.~\ref{fig:CR_vs_dist} the results for J1823-3021G only as the behavior is similar for all the targets.
    
We observe that in the regime $\Tfft < \TfftBin$ (top plots of Fig.~\ref{fig:CR_vs_dist}), there is not a loss of CR associated with the introduced offset.

More interesting are the other two regimes, where the validity of the method is properly tested.
In fact, we observe that, depending on the combination of parameters, a significant reduction of $CR$ occurs if the $\Tfft$ is not chosen properly (see bottom right plot of Fig.~\ref{fig:CR_vs_dist}). 
On the other hand, if we set $\Tfft$ from Eq.~\eqref{eq:condition_coherence_time}, as in the bottom left plot of Fig.~\ref{fig:CR_vs_dist}, we observe a contained loss of significance that increases with larger offsets, as one can expect. 
We stress that the loss that we are showing in Fig.~\ref{fig:CR_vs_dist} might be overestimated because is evaluated with steps of $\sigma_{x_i}$.

\par Wrapping up, we can conclude that tuning the coherence time according to the uncertainties on the orbital parameters allows for a single Doppler correction performed with the ephemeris central values. In this way, we are already sensitive to CW signals without the need for a grid in the orbital parameters, thus saving computing time at a reasonable sensitivity loss. \section{Results}
\label{sec:results}
\begin{figure*}[!htpb]
        \centering
        \includegraphics[width=0.43\textwidth]{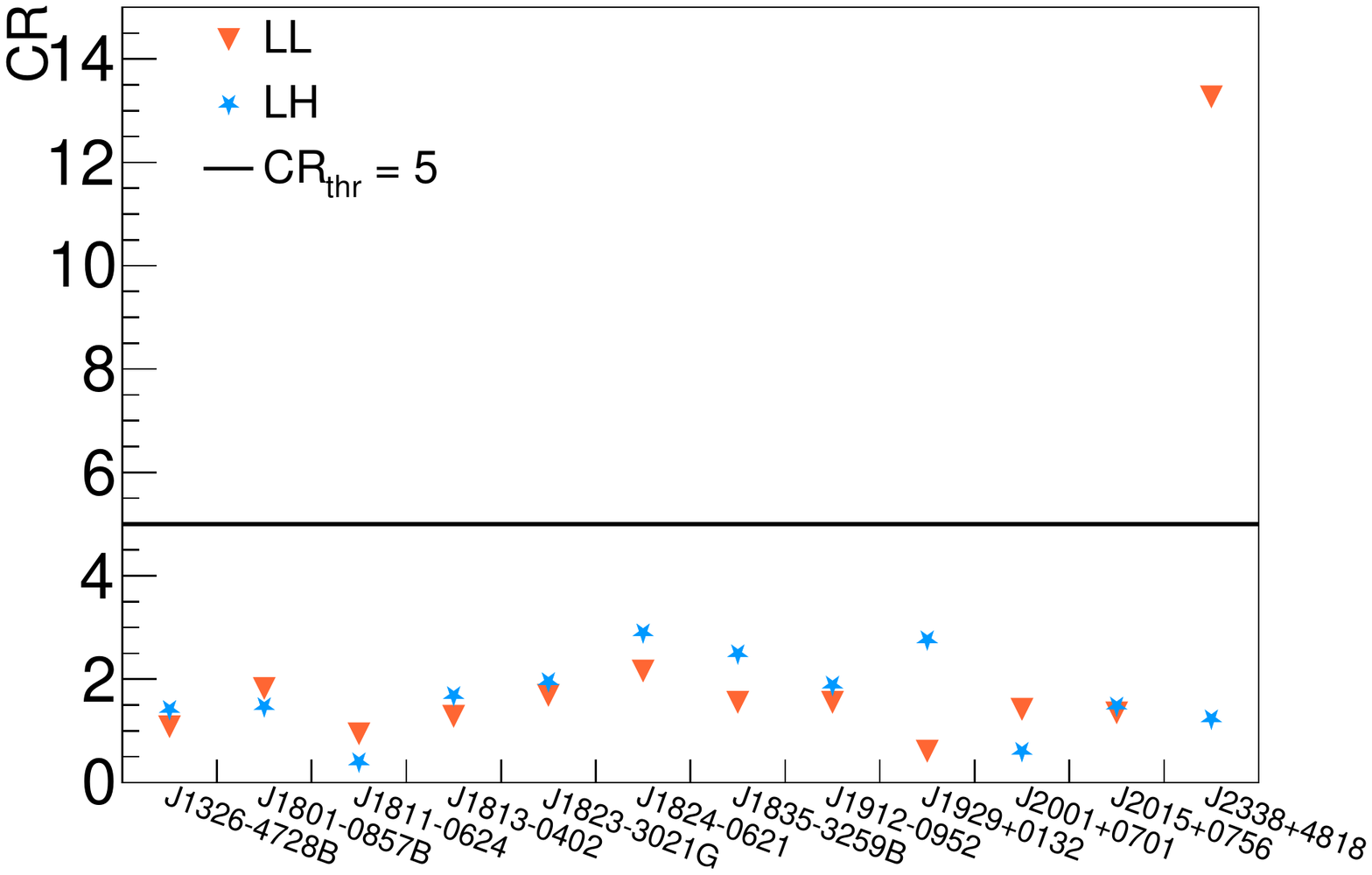}
        \includegraphics[width=0.43\textwidth]{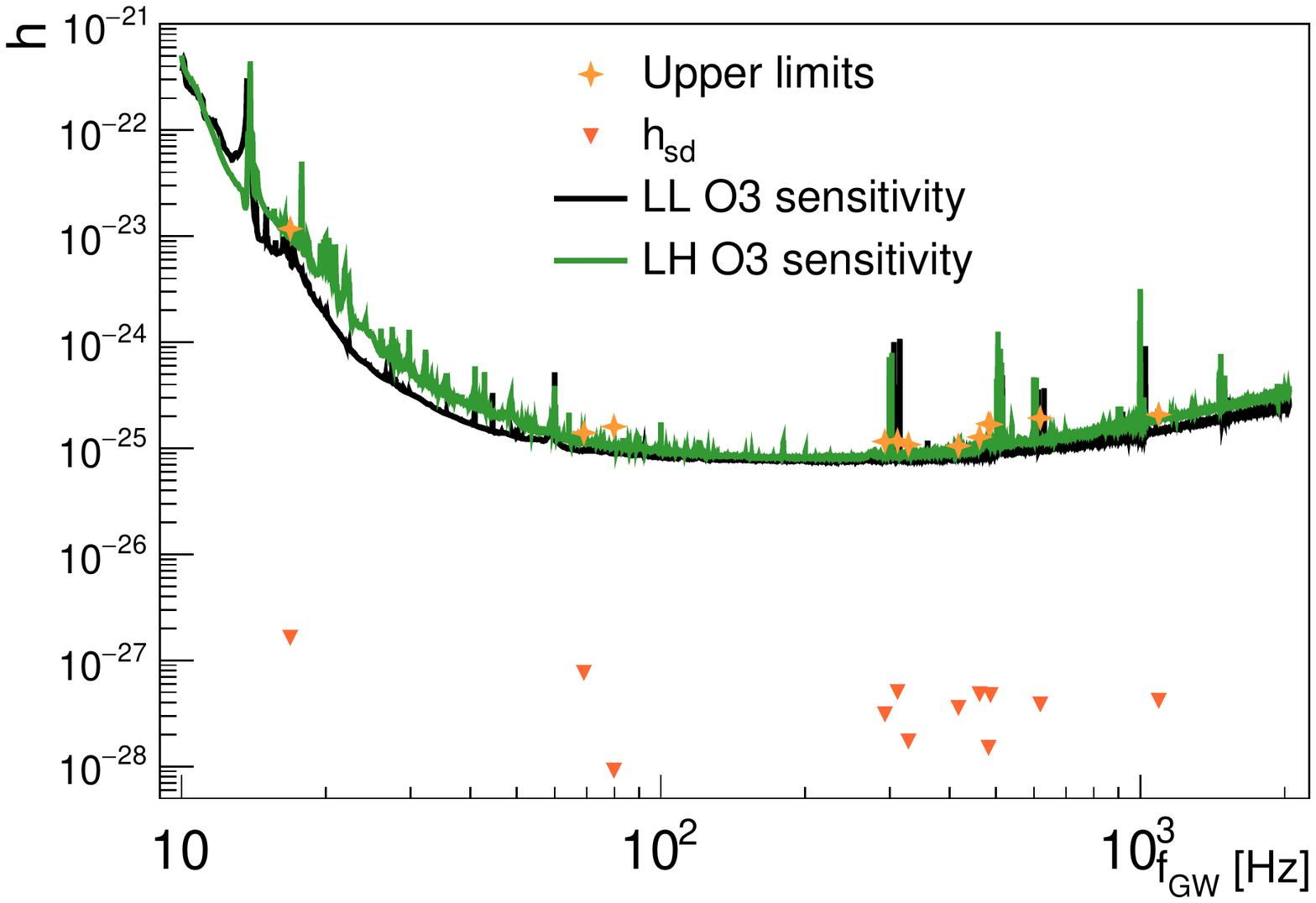}
        \caption{(Left: critical ratio obtained for each of the targets reported in Table~\ref{tab:candidates} after performing the Doppler correction at the ephemeris central values. The AV refers to the results obtained with Virgo data. One outlier has been identified for pulsar J2338+4818 that has been discarded (see text for details). Right: ULs on the strain amplitude (ocher diamonds) at the 95\% CL emitted by the analyzed targets (see text). The ULs are compared with the spin-down limits (see Eq.~\eqref{eq:hsd}) of the targets (orange triangles), and the estimated sensitivity curves from Eq.~\eqref{eq:h0min}.}
        \label{fig:results-UL}
    \end{figure*}

    \begin{figure*}[btp]
        \centering
        \includegraphics[scale=0.3]{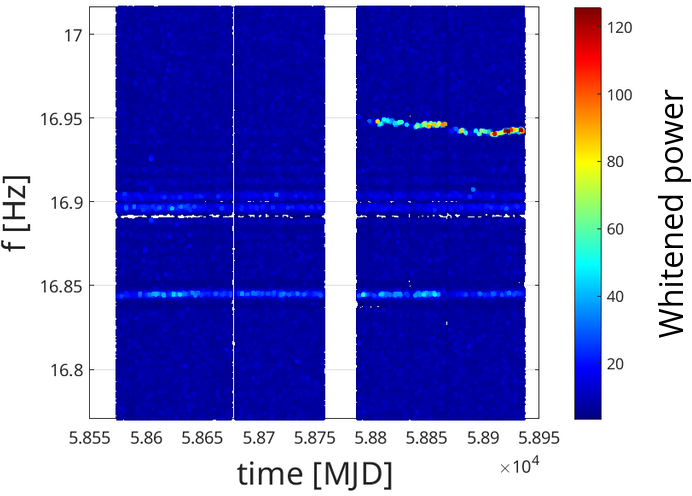}
        \includegraphics[scale=0.3]{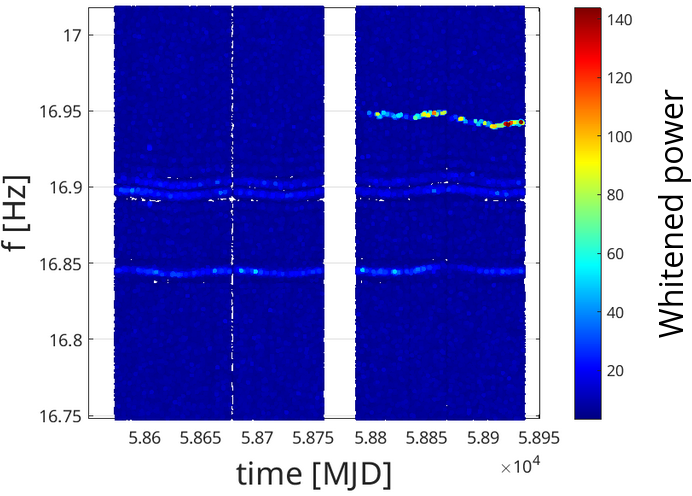}
        \caption{Peakmaps of the LL detector before (\textit{left}) and after (\textit{right}) Doppler correction for the pulsar J2338+4818. The search highlighted an outlier (see Fig.~\ref{fig:results-UL}) that is likely to be of instrumental origin, due to a persistent line at $f\sim16.85$~Hz.}
       \label{fig:PM_around_outlier}
    \end{figure*}
In this section, we apply the search method to both LL and LIGO Hanford (LH) O3 real data~\cite{Abbott_2023}. Virgo and KAGRA data are not included systematically in the analysis as their sensitivities are, almost for all the bands, worse. We considered Virgo data only for pulsar J2338+4818, since at low frequencies its sensitivity is comparable with that of LH.

\par Based on the test with the SIs reported in Sec.~\ref{subsec:test_with_SI}, for the targets considered here, we can safely perform a single Doppler correction at the ephemeris central values.
\par The threshold to identify potentially interesting outliers has been set at $CR_{thr} \sim 2.79$.
In this work, we found one outlier with $CR>CR_{thr}$, see Fig.~\ref{fig:results-UL}, for pulsar J2338+4818, for which the astrophysical origin has been confidently discarded. The outlier is due to a clear persistent excess power, linked to a calibration line. This can be seen comparing the peakmap with and without the Doppler correction, see Fig.~\ref{fig:PM_around_outlier}.

The procedure to select a candidate is as in Sec.~\ref{sec:method} but since no CW-related one has been found, we computed the upper limits (ULs) for all the pulsars through an injection campaign in LL and LH data.
    
\par The calculation proceeds similarly as in~\cite{Abbott_2019} with the difference that we incorporate the binary modulation. We inject one by one a set of 100 SIs into the LL and LH full O3 data with a frequency $\pm 5$ bins away from the target's one, and we start from a strain amplitude of $H_0 = h_{sens}/2$. The factor 1/2 accounts for the fact that the UL for a specific pulsar can be underneath the sensitivity curve as Eq.~\eqref{eq:h0min} is averaged over all sky position and CW parameters~\cite{Astone:2014esa}. $\Psi$ and $\cos\iota$ values are randomly drawn from a flat distributions within $[-45,\,45]$ and $[-1,1]$, respectively.

Following the test method (see Sec.~\ref{sec:method} and App.~\ref{app:other_par_test}), we inject each SI with random orbital parameters taken from Gaussians around the measured ephemeris values, while the Doppler correction is performed with the central ones. As previously mentioned, the frequency of the SIs is taken to be outside the signal region (i.e., the region where the candidate's peak is selected) to avoid overlap. If less than 95\% of the signals are producing $CR\geq CR_{thr}$~\footnote{In the case of the outlier in LL data, the UL is calculated with the measured $CR$.}, another set of SIs is injected with a strain increased by $h_{sens}/10$. This process is repeated till the condition is satisfied, and, in that case, the UL for the single detector is found to be
    \begin{equation}
        h_{\mathrm{UL}} ^{idet} \approx 1.31 \cdot H_{0\,\mathrm{UL}}^{idet},
    \end{equation}
being $H_{0\,\mathrm{UL}}^{idet}$ the $H_0$ single detector's UL, while the factor 1.31 comes from averaging $h_0$ as a function of $H_0$ (i.e., inversion of Eq.~\eqref{eq:Capital_H}) over a flat $\cos\iota$ distribution. Then, we choose the highest UL between the two detectors as the UL for the single target (see Table~\ref{tab:candidates} for our estimations). Our ULs are presented on the right of Fig.~\ref{fig:results-UL} and are compared with the expected sensitivity curves of three detectors involved in the analysis and the spin-down limit calculated as
{\small\begin{equation}
   h_{\rm sd} =  \frac{1}{D} \sqrt{\frac{5 G}{4 c^3}  I_{zz} \bigg\vert\frac{\dot{f}_{0}}{f_{0}}\bigg\vert }
   \label{eq:hsd}
\end{equation}}
for a NS with moment of inertia $I_{zz}$~=~$10^{38}$~kg~m$^2$. \section{Conclusion}
\label{sec:conclusion}
In this paper, we have described a new semicoherent method that is highly computationally efficient to search for CW signals from known pulsars in binary systems.~These targets can be more challenging to study as the binary Doppler correction can rapidly change within the binary orbital parameter uncertainties, potentially reducing the SNR.
\par The efficiency of the method here shown is a combination of two aspects: a custom optimization of the coherence time to reduce the power loss; and a single Doppler correction to assess the presence of a signal, drastically reducing the needed computational cost.
\par Independently on the method, our coherence time calculation can be of crucial relevance for future targeted searches where NS in binary systems are involved. With our calculation, we have given a quantitative estimate to understand whether a pulsar can be safely studied with a single-template fully coherent approach or not.
\par The coherence time is evaluated with the ephemeris binary orbital parameters and their uncertainties, requiring that the variation of the Doppler effect must be confined in half of a frequency bin. Applying this method to thirteen targets from the ATNF catalog version 1.70, we found coherence times that span several orders of magnitudes, from $\sim 0.4$~d, up to $\sim15$~yr.
\par Before applying the method, we tested its reliability with injections in simulated data of realistic duration. Additional tests confirmed how crucial the choice of the coherence time length can be.
\par In order to achieve better sensitivities longer coherence times are needed. We have already foreseen crucial improvements from the software side that will overcome the sidereal day that is now limiting the present implementation.
\par By applying the method to look for a real CW coming from the selected targets, we have found one outlier that we assessed to be of instrumental origin. We then computed upper limits on the gravitational wave strain amplitude of these targets for the very first time to our knowledge, through the injections of fake signals into real data. \section*{Acknowledgment}
We would like to thank the PRD reviewer for fruitful comments that improved this work.\\
This research has made use of data or software obtained from the Gravitational Wave Open Science Center~\cite{gwosc}, a service of the LIGO Scientific Collaboration, the Virgo Collaboration, and KAGRA. This material is based upon work supported by NSF's LIGO Laboratory which is a major facility fully funded by the National Science Foundation, as well as the Science and Technology Facilities Council (STFC) of the United Kingdom, the Max-Planck-Society (MPS), and the State of Niedersachsen/Germany for support of the construction of Advanced LIGO and construction and operation of the GEO600 detector. Additional support for Advanced LIGO was provided by the Australian Research Council. Virgo is funded, through the European Gravitational Observatory (EGO), by the French Centre National de Recherche Scientifique (CNRS), the Italian Istituto Nazionale di Fisica Nucleare (INFN) and the Dutch Nikhef, with contributions by institutions from Belgium, Germany, Greece, Hungary, Ireland, Japan, Monaco, Poland, Portugal, Spain. KAGRA is supported by Ministry of Education, Culture, Sports, Science and Technology (MEXT), Japan Society for the Promotion of Science (JSPS) in Japan; National Research Foundation (NRF) and Ministry of Science and ICT (MSIT) in Korea; Academia Sinica (AS) and National Science and Technology Council (NSTC) in Taiwan.\\
We would like to acknowledge CNAF for providing the computational resources.\\
We would like to also acknowledge the ATNF Pulsar Catalog, Manchester \textit{et al.}~\cite{Manchester:2004bp}, see Ref.~\cite{ATNF_link} for updated versions. \appendix

\section{BINARY DOPPLER VARIATION\label{app:doppl_var}}
As described in Sec.~\ref{sec:method}, our coherence time calculation relies on the variation of the orbital Doppler modulation expressed by Eq.~\eqref{eq:delta_doppl}. Here we detail its calculation.\\
As mentioned in Sec.~\ref{sec:pulsars}, ephemerides are given with either the parameters ($a_p,\, P_{\mathrm{orb}},\,e,\,\omega,\,t_p$), or ($a_p,\, P_{\mathrm{orb}},\,\eta,\,\kappa,\,t_{\mathrm{asc}}$).\\
Let us now focus singularly on each of the cases.

\subsection{Low eccentricity regime}
We use the low eccentricity approximation (see Eq.~\eqref{eq:R_c_low_ecc}) whenever $\eta$, $\kappa$ and $t_{\rm asc}$ are given or the eccentricity is lower than 0.1 \cite{PhysRevD.91.102003}. Since in the former scenario $\eta$, $\kappa$ and $t_{\rm asc}$ are calculated using other parameters [see Eqs.~\eqref{eq:kappa-eta}-\eqref{eq:psi-tasc}], the Doppler variation will have to keep that into account.
\par Let us start from the common term given by the derivative of Eq.~\eqref{eq:low_ecc_R_c_dot} with respect to $a_p$
    {\small
    \begin{equation}
        \frac{\partial \Dot{R}/c}{\partial a_p} = \frac{1}{a_p}\cdot\frac{\Dot{R}}{c}\,.
    \end{equation}}

Considering the scenario where the lagrangian parameters are given, the other contributions are calculated as
    {\small
    \begin{gather}
        \frac{\partial \Dot{R}/c}{\partial P_{\mathrm{orb}}} = -\frac{1}{P_{\mathrm{orb}}} \bigg( \frac{\Dot{R}}{c} \nonumber\\ 
        - \,a_p \,\Omega \, \psi \, (\sin\psi + 2\kappa\,\sin2\psi - 2\eta\,\cos2\psi) \bigg)\,,\\
        \frac{\partial \Dot{R}/c}{\partial t_{\mathrm{asc}}} = a_p\, \Omega^2 \left( \sin\,\psi + 2\kappa\,\sin\,2\psi - 2\eta\,\cos\,2\psi\right)\,,\\
        \frac{\partial \Dot{R}/c}{\partial \eta} = a_p \, \Omega \, \sin\,2\psi\,,\\
        \frac{\partial \Dot{R}/c}{\partial \kappa} = a_p \, \Omega \, \cos\,2\psi\,.
    \end{gather}}
The contributions are calculated differently if $e,\,\omega,\,t_p$ are given
    {\small
    \begin{gather}
        \frac{\partial \Dot{R}/c}{\partial P_{\mathrm{orb}}} = - \frac{1}{P_{\mathrm{orb}}} \bigg(\frac{\Dot{R}}{c} \nonumber\\
        +\,a_p \Omega (\omega-\psi) (\sin\psi + 2\kappa\sin2\psi - 2\eta\cos2\psi) \bigg)\,,\\
        \frac{\partial \Dot{R}/c}{\partial t_p} = \frac{\partial \Dot{R}/c}{\partial t_{\mathrm{asc}}}\,,\\
        \frac{\partial \Dot{R}/c}{\partial e} = a_p\, \Omega \, (\cos2\psi\,\cos\omega + \sin2\psi\,\sin\omega)\,,\\
        \frac{\partial \Dot{R}/c}{\partial \omega} = - a_p \Omega\, (\sin\psi + 2\kappa\,\sin2\psi - 2\eta\,\cos2\psi)\,.
    \end{gather}}

\subsection{High eccentricity regime}
On the other hand, the Doppler variation in the high eccentricity regime is calculated as follows
{\small
\begin{gather}
    \frac{\partial \Dot{R}/c}{\partial E} = \frac{a_p \Omega \,  ( \sin\omega\,(e-\cos E) - \cos E\sin\omega \sqrt{1-e^2} ) }{(1-e\cos E)^2}\\
    \frac{\partial \Dot{R}/c}{\partial P_{\mathrm{orb}}} = -\frac{\Omega}{P_{\mathrm{orb}}} \, \frac{\partial \Dot{R}/c}{\partial E} \, \frac{t-t_p}{1-e\cos E}\,,\\
    \frac{\partial \Dot{R}/c}{\partial t_p} = -\frac{\Omega}{1-e\cos E} \, \frac{\partial \Dot{R}/c}{\partial E}\,,\\
\end{gather}

    \begin{widetext}
        \centering
        \begin{gather}
        \frac{\partial \Dot{R}/c}{\partial e} = \frac{\sin E}{1-e \cos E} \frac{\partial \Dot{R}/c}{\partial E} \notag\\ \frac{a_p\Omega}{(1-e \cos E)^2}\bigg( \cos E \left(\frac{\cos E-e}{\sqrt{1-e^2}}\cos\omega - \sin\omega \,\sin E\right) + \sin^2 E \,\frac{\sin\omega\,(e-\cos E) - \cos\omega\,\sin E\,\sqrt{1-e^2}}{\sin E - \cos E (E - \Omega\,(t-t_p))}  \bigg)\,,
        \end{gather}
    \end{widetext}

\begin{equation}
    \frac{\partial \Dot{R}/c}{\partial \omega} = \frac{a_p\Omega}{1-e\cos E}\,(\sin\omega \cos E \,\sqrt{1-e^2} + \cos\omega\,\sin E)\,.
\end{equation}}
Equation~\eqref{eq:delta_doppl} is evaluated taking the sum of the terms listed in this appendix, weighted by the uncertainty on each parameter.

\section{UPPER BOUND TO THE COHERENCE TIME\label{app:Tfft_max}}
\begin{figure*}[htb]
        \centering
        \includegraphics[scale=0.33]{APS_P1.pdf}
        \includegraphics[scale=0.33]{APS_SI.pdf}
        \includegraphics[scale=0.33]{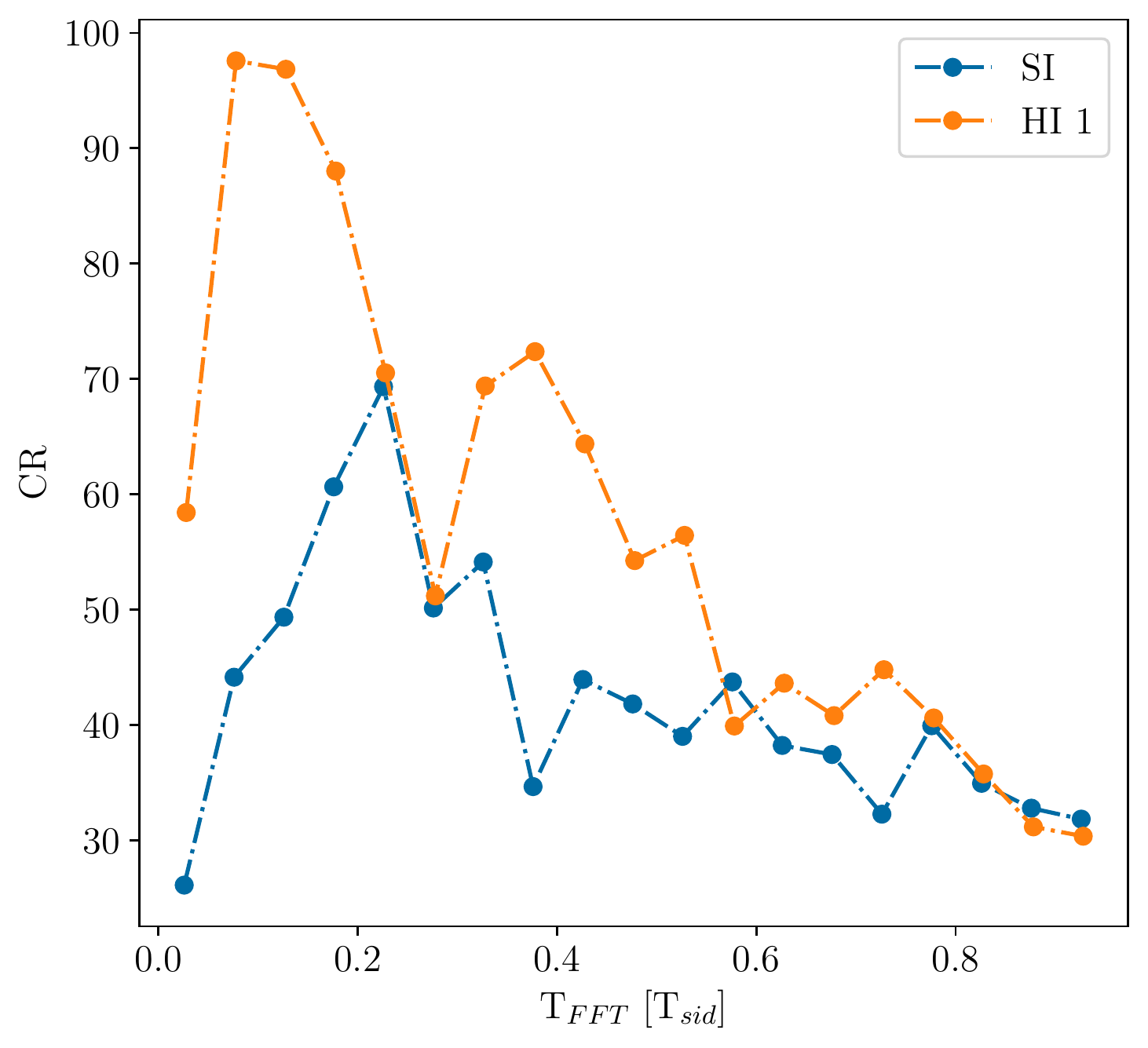}
        \caption{Power spectra of the HI 1 (left) and a SI (middle) for which the 5 harmonics of the signals are visible, see Table~\ref{tab:inj_par} for the injection parameters. (right) CR as a function of the coherence time. Too high values tend to disperse the signal power, lowering the significance of the candidate.}
       \label{fig:CR_vs_Tfft}
    \end{figure*}
    
The current implementation of the method presented in Sec.~\ref{sec:method} does not account for the Earth's rotational Doppler modulation. Due to the harmonic components of the beam pattern function \cite{Jaranowski:1999pd}, the power of the monochromatic CW is split into up to five components at $f_{0} \pm k/T_{sid}$, with $k=0,\>1,\>2$, limiting the coherence time of a search if not correctly kept into account.\\
Since the relative height of the harmonics is not constant but rather depends on the detector and source locations~\cite{PhysRevD.58.063001}, the splitting is not uniform and rules the number of peaks found above the noise level.\\
We set the highest coherence time of the search by studying the worst-case scenario, where the splitting is almost even among the harmonics. In the top part of Fig.~\ref{fig:CR_vs_Tfft} we show the power spectrum of the \textit{hardware injection}\footnote{CW injected through the detector hardware for testing purposes, see~\cite{gwosc} for the injection parameters.} (HI) 1 and an isolated NS SI. We report the injections' parameters in Table~\ref{tab:inj_par}, where the harmonics are visible after removing all the Doppler modulation except the sidereal one. We analyse both the signals with different coherence time in $\Tfft \in [0.02\,T_{sid},\, T_{sid}]$ with a step of $0.05 \,T_{ sid}$. The bottom plot in Fig.~\ref{fig:CR_vs_Tfft} shows the CR obtained with each of the $\Tfft$ values. We observe a decreasing trend for both the SI and HI1 that starts around $\Tfft\sim 0.2\, T_{sid}$ confirming this effect can significantly reduce the detection chances if not properly accounted for. For this reason, we set $0.2\, T_{sid}$ as our highest coherence time. 
    \begin{table}[ht]
        \centering
        \renewcommand*{\arraystretch}{1.1}
        \begin{tabular}{ccc}
             \hline\hline
             & HI 1 & SI \\
             \hline
             $\alpha$ &  \ang{37.39}  & \ang{121.98}\\ 
             $\delta$ &  -\ang{29.45}  & -\ang{21.49}\\
             $f_{0}$ [Hz] &  848.937  &  101.217\\
             $\Dot{f}_{0}$  [Hz/s] &  -3$\cdot10^{-10}$  & 0\\
             $h_0$  & 1.68$\cdot10^{-24}$ & 2.74$\cdot10^{-25}$ \\
             $\cos\iota$ & 0.46 & 0.34\\
             $\Psi$ & \ang{20.40}  & -\ang{28.60}\\
             \hline\hline
        \end{tabular}
        \caption{Injection parameters of the signals shown in Fig.~\ref{fig:CR_vs_Tfft}. The frequency reference time is given at $t_{\rm ref}$ 58574~MJD.}
        \label{tab:inj_par}
    \end{table}

\section{COHERENCE TIME OF ATNF PULSARS IN BINARY SYSTEMS\label{app:Tfft_ATNF_binaries}}
We selected from the ATNF catalog the pulsars in binary systems using the python package \texttt{psrqpy}~\cite{psrqpy} applying the same criteria described in Sec.~\ref{sec:pulsars} without the requirement on the frequency reference time.
In Fig.~\ref{fig:Tfft_ATNF} we show the histogrammed $\TfftBin$ that we obtained from these targets with Eq.~\eqref{eq:condition_coherence_time}. Our estimation highlights a non-null population of pulsars (31 out of 235) with $\TfftBin<T_{\rm FFT}^{max}$.

\begin{figure}[ht]
    \centering
    \includegraphics[width=0.42\textwidth]{Tfft_ATNF.pdf}
    \caption{Histogrammed $\TfftBin$ of 235 ATNF binary pulsars calculated with Eq.~\eqref{eq:condition_coherence_time}.}
    \label{fig:Tfft_ATNF}
\end{figure}

\section{UNCERTAINTIES ON SKY LOCATION\label{app:Tfft_err_sky_pos}}
Although in this work we focused on uncertainties on the orbital parameters, the sky position indetermination can play, in principle, a non-negligible role. Here we present an ``equivalent'' coherence time calculation based on~\cite{Astone:2014esa}.

\begin{figure}[ht]
    \centering
    \includegraphics[width=0.45\textwidth]{Tfft_err_pos_vs_Tfft.pdf}
    \caption{Coherence time given by uncertainties on sky location (Eq.~\eqref{eq:Tfft_err_pos}) plotted against the $\Tfft$ given by the uncertainties on binary parameters (Eq.~\eqref{eq:condition_coherence_time}).}
    \label{fig:Tfft_err_pos}
\end{figure}

\par Following Eqs.~(38)-(42) of~\cite{Astone:2014esa} we can link the resolution in the sky parameters\footnote{We make use of the ecliptic coordinates $\lambda$ and $\beta$.} to the $\Tfft$
{\small\begin{gather}
    \Tfft^{\rm sky}(\delta\lambda) \approx \frac{10^4}{f_{0}\,\delta\lambda\,\cos\beta}\,, \quad
    \Tfft^{\rm sky}(\delta\beta) \approx \frac{10^4}{f_{0}\,\delta\beta\,\sin\beta}\,,
    \label{eq:Tfft_err_pos}
\end{gather}}
where $10^4$ is approximately the inverse ratio of the Earth's orbital velocity and $c$. Here we are interpreting $\delta\lambda$ and $\delta\beta$ as the uncertainties on the sky location given by astronomers. 

The comparison of $min\left(\Tfft^{\rm sky}(\delta\lambda),\Tfft^{\rm sky}(\delta\beta)\right)$ (i.e., the worst-case scenario between the two) with the $\TfftBin$ from Eq.~\eqref{eq:condition_coherence_time} allows to determine whether the binary orbital uncertainties are dominating over the others.
For this reason, considering the values reported in Table~\ref{tab:err_pos} for each of our targets, in Fig.~\ref{fig:Tfft_err_pos} we plot the coherence time calculated with the two sets of uncertainties one against the other. As we can see, $\Tfft^{\rm sky}$ is greater than the other estimation by a factor of $O(10^{1-3})$ depending on the considered target.

    \begin{table}[ht]
        \centering
        \renewcommand*{\arraystretch}{1.1}
        \setlength{\tabcolsep}{7pt}
        \begin{tabular}{cccc}
             \hline\hline
             Name & $\beta$ [$\degree$]& $\delta\beta$ [$\degree$] & $\delta\lambda$ [$\degree$]\\ \hline
        J1326-4728B & -35.23433 & 1.3$\cdot 10^{-5}$ & 1.5$\cdot 10^{-5}$\\
        J1701-3006G & -7.3218 & 1.1$\cdot 10^{-2}$ & 4$\cdot 10^{-4}$\\
        J1801-0857B & 14.479409 & 6$\cdot 10^{-6}$ & 1.7$\cdot 10^{-6}$\\
        J1811-0624  & 17.000105 & 8$\cdot 10^{-6}$ & $10^{-6}$\\
        J1813-0402  & 19.3503669 & 3$\cdot 10^{-7}$ & $10^{-7}$\\
        J1823-3021G & -7.02626 & 1.4$\cdot 10^{-5}$ & 3$\cdot 10^{-6}$\\
        J1824-0621  & 16.9408957 & 4$\cdot 10^{-7}$ & $10^{-7}$\\
        J1835-3259B & -9.786669 & 3$\cdot 10^{-6}$ & 5$\cdot 10^{-7}$\\
        J1912-0952  & 12.424092 & 4$\cdot 10^{-6}$ & 1.8$\cdot 10^{-6}$\\
        J1929+0132  & 23.0986161 & 3$\cdot 10^{-7}$ & 1.5$\cdot 10^{-7}$\\
        J2001+0701  & 26.9071496 & 3$\cdot 10^{-7}$ & 2$\cdot 10^{-7}$\\
        J2015+0756  & 26.9623559 & 4$\cdot 10^{-7}$ & 3$\cdot 10^{-7}$\\
        J2338+4818  & 45.264334 & 4$\cdot 10^{-6}$ & 4$\cdot 10^{-5}$\\\hline\hline
        \end{tabular}
        \caption{Ecliptic latitude $\beta$ and uncertainties on sky location used to evaluate Eq.~\eqref{eq:Tfft_err_pos} for each of the targets in Table~\ref{tab:candidates}.}
        \label{tab:err_pos}
    \end{table}

We leave as a future work the estimation of the coherence time based on the thorough variation of the Doppler modulation. It would be eventually needed wherever the two evaluations [Eqs.~\eqref{eq:condition_coherence_time}-\eqref{eq:Tfft_err_pos}] are comparable. As for now, we would suggest considering a gridded approach separately on the two sets of parameters or rather discarding the target.

\section{\texorpdfstring{$CR$}{CR} NOISE DISTRIBUTION\label{app:noise_distr}}
Here we detail the FAP calculation at a given $CR_{\rm thr}$.
\par In the absence of signals and under the Gaussian noise hypothesis, the peakmap consists of frequency bins selected according to a binomial distribution~\cite{Astone:2014esa} with probability equal to $p_0$ (see Eq.~\eqref{eq:FAP_single_Tfft}) and number of trials $2N_{\rm FFT}$, with the factor of 2 due to the half-interlaced FFTs.
\par We are interested in the distribution followed by the maximum $CR$ over $N_{\rm bin} = 3$ bins since we select the loudest frequency bin within $f_{\rm 0}\pm 1.5\> \delta f$, see Sec.~\ref{sec:method}. As a result, the FAP can be estimated through the relation
{\small\begin{equation}
    \mathrm{FAP} (CR_{\rm thr}) = 1-\left(cdf[B_N(x_{\rm thr},\,2N_{\rm FFT},\,p_0)]\right)^{N_{\rm bin}}\,
    \label{eq:FAP}
\end{equation}}
with $B_N$ indicating the binomial distribution described above, and $cdf$ its cumulative distribution function evaluated at
{\small\begin{equation}
    x_{thr}(CR_{\rm thr}) = CR_{\rm thr}\cdot\sigma + \mu\,
\end{equation}}
that is Eq.~\eqref{eq:CR} inverted. The $CR_{\rm thr}$ corresponding to a fixed FAP value can then be estimated through root-finding algorithms.
\par We show in Fig.~\ref{fig:CR_noise_distr} between the FAP estimated through Eq.~\eqref{eq:FAP} and simulations.

\begin{figure}[tb]
    \centering
    \includegraphics[width=0.42\textwidth]{CR_noise_distr.pdf}
    \caption{FAP evaluated with Eq.~\eqref{eq:FAP} (solid blue line) compared with simulated gapless, Gaussian noise-only data (orange dots) for $T_{\rm obs} = 1$~yr and $\Tfft=17236$~s.}
    \label{fig:CR_noise_distr}
\end{figure}

\section{Additional tests\label{app:other_par_test}}
As a further test, we performed a similar study to what is described in Sec.~\ref{sec:method} to check that our method is independent on $\cos\iota$ and $\Psi$.

We injected 500 fake signals resembling J1823-3021G with uncertainties on binary parameters enlarged by a factor of 20. Each injection has $\cos\iota$ and $\Psi$ randomly drawn from flat distributions within $[-1,\,1]$ and $[-45,\,45]$, respectively. Injections are made at $2\,h_{\rm sens}$ to avoid, in the $\cos\iota\sim0$ regime, a $CR$ comparable to noise (the minimum $CR_{\rm OFF}$ obtained during this test is above 10).

For each injection, we perform two Doppler corrections to evaluate $r(d)$: one using the injected parameters, the parameters of the second correction are taken randomly from Gaussian distributions built around the injected parameters and with standard deviation equal to their uncertainties.
Then, we calculate $r(d)$ by taking the ratio of the two $CR$s calculated after the corrections.

The outcome of this test is shown in Fig.~\ref{fig:iota_psi_rnd_inj}.
We observe the same trend as in Fig.~\ref{fig:CR_vs_dist}, without highlighting any clear dependence on $\cos\iota$ or $\Psi$.

\begin{figure*}[tb]
    \centering
    \includegraphics[width=0.4\textwidth]{CR_random_injections_err_larger_iota.pdf}
    \includegraphics[width=0.4\textwidth]{CR_random_injections_err_larger_psi.pdf}
    \caption{$r(d)$ for 500 injected signals with $\cos\iota$ and $\Psi$ randomly drawn from flat distributions within $[-1,\,1]$ and $[-45,\,45]$, respectively. $CR_{\rm OFF}$ is calculated for a random set of binary parameters drawn from Gaussian distributions built around the injected ones and with a standard deviation equal to their uncertainties. The two plots are identical with the color code indicating $\cos\iota$ (\textit{left}) and $\Psi$ (\textit{right}) for each injection.}
    \label{fig:iota_psi_rnd_inj}
\end{figure*}

\bibliographystyle{bibstyle.bst}
\bibliography{refs}

\end{document}